\documentclass[11pt,a4paper]{article}
\pdfoutput=1
\usepackage{jcappub}
\usepackage{graphicx}
\usepackage{longtable}
\usepackage{amsmath}
\usepackage{amssymb}
\usepackage{subfigure}
\usepackage{hyperref}
\usepackage{dcolumn}
\usepackage{multirow}
\usepackage{bm}
\usepackage{color}

\newcommand{\be}{\begin{eqnarray}}
\newcommand{\ee}{\end{eqnarray}}
\newcommand{\mpl}{{M_{\rm {pl}}}}

\newcommand{\dd}{\, {\rm d}}

\def\bea{\begin{eqnarray}}
\def\eea{\end{eqnarray}}

\newcommand{\rv}{r_{\rm v}}

\allowdisplaybreaks
\begin{document}
\title{Testing Gravity Using Galaxy Clusters: New Constraints on Beyond Horndeski Theories}
\author{Jeremy Sakstein,}
\emailAdd{jeremy.sakstein@port.ac.uk}
\author{Harry Wilcox,}
\emailAdd{harry.wilcox@port.ac.uk}
\author{David Bacon,}
\emailAdd{david.bacon@port.ac.uk}
\author{Kazuya Koyama,}
\emailAdd{kazuya.koyama@port.ac.uk}
\author{and Robert C. Nichol}
\emailAdd{bob.nichol@port.ac.uk}
\affiliation{Institute of Cosmology and Gravitation,
University of Portsmouth, Portsmouth PO1 3FX, UK}

\abstract{The Beyond Horndeski class of alternative gravity theories allow for Self-accelerating de-Sitter cosmologies with no need for a cosmological constant. This makes them viable alternatives to $\Lambda$CDM and so testing their small-scale predictions against General Relativity is of paramount importance. These theories generically predict deviations in both the Newtonian force law and the gravitational lensing of light inside extended objects. Therefore, by simultaneously fitting the X-ray and lensing profiles of galaxy clusters new constraints can be obtained. In this work, we apply this methodology to the stacked profiles of 58 high-redshift ($ 0.1<z<1.2$) clusters using X-ray surface brightness profiles from the XMM Cluster Survey and weak lensing profiles from CFHTLenS. By performing a multi-parameter Markov chain Monte Carlo analysis, we are able to place new constraints on the parameters governing deviations from Newton's law $\Upsilon_{1}=-0.11^{+0.93}_{-0.67}$ and light bending $\Upsilon_{2}=-0.22^{+1.22}_{-1.19}$. Both constraints are consistent with General Relativity, for which $\Upsilon_{1}=\Upsilon_{2}=0$. We present here the first observational constraints on $\Upsilon_{2}$, as well as the first extragalactic measurement of both parameters.}
\maketitle

\section{Introduction}

The elusive nature of dark energy \cite{Copeland:2006wr} has prompted considerable research effort into alternative theories of gravity as a potential driving mechanism for the acceleration of the cosmic expansion (see \cite{Clifton:2011jh, Koyama:2015vza}), the most common and well-studied being scalar field modifications. On smaller scales, consistency with Solar System and other tests of General Relativity (GR) is achieved using \textit{screening mechanisms} \cite{Joyce:2014kja}. These are non-linear features of the theory that ensure that any additional degrees of freedom decouple in the Solar System despite the fact that they are relevant on cosmological scales. Typically, these come in two varieties: those that screen by suppressing the scalar charge such as the chameleon \cite{Khoury:2003rn} and symmetron \cite{Hinterbichler:2010es} mechanisms, and those which screen by suppressing the scalar field gradient, such as the K-mouflage \cite{Babichev:2009ee, Barreira:2015aea, Brax:2015lra} and Vainshtein \cite{Vainshtein:1972sx,Nicolis:2008in,Koyama:2013paa,Kimura:2011dc} mechanisms\footnote{See \cite{Hui:2009kc} for a discussion of the differences between these mechanisms.}. 

This paper is concerned with the Vainshtein mechanism, which is ubiquitous in scalar-tensor alternative gravity theories. First seen in the non-relativistic limit of Lorentz-invariant massive gravity theories \cite{Vainshtein:1972sx}, the mechanism was re-discovered in the context of DGP braneworld models \cite{Dvali:2000hr, Deffayet:2001uk, Luty:2003vm}, whose decoupling contains a higher-order derivative interaction known as the \textit{cubic galileon} \cite{Nicolis:2008in} (so called because it involves three fields). Despite its higher-order nature, a special \textit{galilean} symmetry ensures that the field equations are second-order and therefore the Ostrogradski ghost instability is avoided. A natural question to ask is whether one can find other healthy higher-derivative interactions, and the answer is indeed yes; one can also find a quartic and quintic galileon\footnote{In four dimensions. There are more (fewer) possibilities in higher (lower) dimensions.} \cite{Nicolis:2008in}. Away from the decoupling limit, it is necessary to add couplings of the scalar to curvature tensors in order to ensure the ghost-free nature of the theory on arbitrary dynamical space-times. In this way, one is led to the Horndeski \cite{Horndeski:1974wa} or covariant galileon \cite{Deffayet:2009wt} theory, the most general scalar-tensor theory that gives rise to manifestly second-order field equations for both the scalar and the metric. 

In Horndeski theories, the Newtonian force profile sourced by an object of mass $M$ is 
\begin{equation}\label{eq:vainhorn}
\frac{\dd \Phi}{\dd r} = \frac{GM}{r^2}\left[1+2\alpha^2\left(\frac{r}{r_{\rm V}}\right)^n\right],
\end{equation}
where $\alpha$ is an $\mathcal{O}(1)$ coupling constant and $r_{\rm V}$ is a length-scale known as the \textit{Vainshtein radius}. The power $n>0$ depends on the specific theory and is equal to $3/2$ in the simplest cubic galileon theories. The first term  is the Newtonian force predicted by GR and the second is a correction due to the scalar fifth-force. On scales smaller than the Vainshtein radius ($r\ll\rv $) this correction is negligible and Newtonian physics is recovered. This is the Vainshtein mechanism. As an example, the Vainshtein radius of the Sun is of order $10^2$ pc \cite{Khoury:2013tda} and so the entire solar system lies well within this. This makes the Vainshtein mechanism difficult to test on small scales\footnote{Indeed, a similar test to the one we perform here yields poor constraints on galileon and non-local models \cite{Barreira:2015fpa}.} although some attempts have been made\footnote{In contrast, small scale tests of chameleon (and similar) mechanisms provide the strongest constraints \cite{Davis:2011qf,Jain:2012tn,Brax:2013uh,Sakstein:2013pda,Vikram:2014uza,Sakstein:2014nfa,Sakstein:2015oqa,Burrage:2014oza,Hamilton:2015zga}.} \cite{Afshordi:2008rd,Hui:2012jb,Hui:2012qt,Falck:2014jwa}. 

Recently, it has been shown that there are healthy extensions of Horndeski's theory \cite{Zumalacarregui:2013pma,Gleyzes:2014dya,Gleyzes:2014qga} deemed \textit{beyond Horndeski} theories. In these theories, the equations of motion are not manifestly second-order in time, but one can always find a combination of equations that is \cite{Deffayet:2015qwa} \footnote{Note that one must be careful when combining Horndeski and Beyond Horndeski theories since the wrong combination can introduce ghosts \cite{Langlois:2015cwa, Langlois:2015skt, Crisostomi:2016tcp}}. In these theories, the Vainshtein mechanism is only partially successful at screening. Defining the perturbed Minkowski line-element via $\dd s^2 = (-1+2\Phi)\dd t^2 + (1+2\Psi)\delta_{ij}\dd x^i\dd x^j$, the metric potentials sourced by an extended object satisfy the same equations as GR outside the object but inside they are modified to
\cite{Kobayashi:2014ida,Koyama:2015oma}
\begin{align}
\frac{\dd\Phi}{\dd r}&= \frac{GM(r)}{r^2}+\frac{\Upsilon_1 G}{4}\frac{\dd^2 M(r)}{\dd r^2}\label{eq:dPhi}\\
\frac{\dd\Psi}{\dd r}&= \frac{GM(r)}{r^2}-\frac{5 \Upsilon_2 G}{4r}\frac{\dd M(r)}{\dd r}\label{eq:dPsi}.
\end{align}
Here $\Upsilon_i$ are dimensionless numbers that depend on the specific theory. In the simplest case of the covariant ($\textrm{G}^3$) quartic galileon one has $\Upsilon_1=\Upsilon_2=(\dot{\phi}/\Lambda)^4$ where $\phi$ is the (dimensionless) cosmological scalar and $\Lambda$ is the mass scale associated with the quartic galileon defined by $\Lambda_4^6=\mpl^2\Lambda^4$. Note that this parameterisation has been chosen to make contact with \cite{Koyama:2015oma,Sakstein:2015aqx,Sakstein:2015zoa,Sakstein:2015aac}. Since $\Phi$ governs the motion of non-relativistic particles, $\Upsilon_1$ parameterises deviations from General Relativity in non-relativistic systems. Typically, the mass of astrophysical objects is more concentrated in the centre (the density decreases radially outwards) and so $\Upsilon_1>0$ ($<0$) corresponds to a weakening (strengthening) of gravity.

Several works \cite{Koyama:2015oma,Saito:2015fza,Sakstein:2015zoa,Sakstein:2015aac, Jain:2015edg} have examined the behaviour of non-relativistic stars in these theories\footnote{Note that \cite{Saito:2015fza} use a different parameter to $\Upsilon_1$. Furthermore, $\Upsilon_1$ is referred to as $\Upsilon$ in \cite{Koyama:2015oma,Saito:2015fza,Sakstein:2015zoa,Sakstein:2015aac}.} and $\Upsilon_1$ is constrained to lie in the interval $-0.51<\Upsilon_1<0.027$ at redshift zero. The lower limit comes from the consistency of the Chandrasekhar mass with the lowest mass white dwarf \cite{Jain:2015edg} and the upper limit comes from the consistency of the minimum mass for hydrogen burning in stars with the lowest mass red dwarf \cite{Sakstein:2015zoa,Sakstein:2015aac}. Note that one requires $\Upsilon_1>-2/3$ at all redshifts in order to have stable spherically static stellar solutions \cite{Saito:2015fza}. $\Upsilon_2$ governs deviations in the motion of light from around non-relativistic objects, and is not presently constrained at any redshift. 

$\Upsilon_i$ are very important parameters because they are related to the parameters appearing in the effective field theory of dark energy (EFT) \cite{Bellini:2014fua,Gleyzes:2014qga} via
\begin{equation}\label{eq:EFT}
\Upsilon_1=\frac{4\alpha_H^2}{c_T^2(1+\alpha_B)-\alpha_H-1}\quad\textrm{and}\quad \Upsilon_2 = \frac{4\alpha_H(\alpha_H-\alpha_B)}{5(c_T^2(1+\alpha_B)-\alpha_H-1)}.
\end{equation}
The first of these equations was derived by \cite{Saito:2015fza}\footnote{Note that $\epsilon\rightarrow-\Upsilon/4$ in their notation.} and we derive the second in appendix \ref{app:U2}. The five parameters that appear in the EFT completely characterise the linear cosmology of beyond Horndeski theories\footnote{Note that, in this sense, beyond Horndeski theories include Horndeski theories as a subset, although modifications of GR of the form studied here only appear on small scales if there is at least one beyond Horndeski term such that $\alpha_H\ne0$.} and therefore constraints on $\Upsilon_i$ directly constrain deviations from GR on cosmological scales. We note for completeness that in this work (like all previous studies) we work with the Jordan frame formulation where matter is minimally coupled to the metric. In this formalism the scalar is coupled to matter via couplings to curvature tensors and not directly via the metric and modifications of gravity hence appear at the level of the equations of motion. It is well-known which region of parameter space is free of pathologies such as laplacian instabilities \cite{DeFelice:2015isa} and this places further restrictions on $\Upsilon_i$.

The aim of this work is to place new constraints on $\Upsilon_1$ by comparing the stacked X-ray and weak lensing profiles of galaxy clusters using a similar method to \cite{Terukina:2013eqa,Wilcox:2015kna} who have used cluster profiles to constrain chameleon models. The intracluster ionised plasma is in hydrostatic equilibrium\footnote{We discuss non-thermal pressure in section \ref{sec:clustersBH}.} and is therefore sensitive to $\Upsilon_1$ through the pressure support equation $\dd\Phi/\dd r=-1/\rho\dd P/\dd r$. The thermal pressure can be directly related to the X-ray surface brightness (see below) and in this work we use the same data and fitting methods as \cite{Wilcox:2015kna}, with X-ray data from the XMM-Newton public archive \cite{Romer:1999qt}. The lensing of light by the dark matter halo is governed by $\Phi+\Psi$ and can hence be used to probe both $\Upsilon_1$ and $\Upsilon_2$. In the case of chameleon models the lensing is unaffected by the modifications of gravity. This is not so for beyond Horndeski theories, and it is this that gives rise to our constraining power. Performing a joint multi-parameter Markov chain Monte Carlo MCMC fit to both the X-ray and lensing profiles, we obtain new constraints on both $\Upsilon_1$ and $\Upsilon_2$ which are consistent with GR. The clusters in our sample span the redshift range $0.1<z<1.2$ with a median value $z_{\rm med}=0.33$, so our constraints apply at this redshift\footnote{One potential caveat is for models where $\Upsilon_i$ vary rapidly over this redshift range. We discuss this and other caveats in section \ref{sec:res}.}. Our constraints are therefore complimentary to those previously obtained at redshift zero, and provide a new hurdle that any successful beyond Horndeski theory must jump in order to be viable.

This paper is organised as follows: In section \ref{sec:clustersBH} we provide a brief introduction to the principles of cluster physics underpinning our test. Since our methodology is identical to \cite{Wilcox:2015kna} we do not repeat it in depth and instead refer the reader there for a more detailed description of the procedure. However, we indicate where the Vainshtein case differs from the chameleon one, which is mostly in the treatment of lensing. In section \ref{sec:res} we present our results and discuss the implications for beyond Horndeski models, as well as necessary caveats. We conclude in section \ref{sec:concs}.

\section{Cluster Properties in Beyond Horndeski Theories}\label{sec:clustersBH}

As discussed in the introduction, here we give a brief overview of cluster physics in beyond Horndeski theories, and our methodology; we refer the interested reader to \cite{Wilcox:2015kna, Wilcox2016} for a more detailed account. We describe the dark matter halo using a Navarro-Frenk-White (NFW) profile \cite{Navarro:1995iw} which has been shown to be appropriate for chameleon gravity models via numerical simulations \cite{Wilcox2016},
\begin{equation}\label{eq:NFW}
\rho_{\rm NFW}(r)=\frac{\rho_{\rm s}}{\frac{r}{r_{\rm s}}(1+\frac{r}{r_{\rm s}})^2}.
\end{equation}
Previous studies of dark matter halos in theories that exhibit Vainshtein screening have found that the dark matter density is well described by the NFW profile \cite{Barreira:2013xea,Barreira:2013eea,Barreira:2014zza}. When the intracluster ionized plasma is in hydrostatic equilibrium one has (using \eqref{eq:dPhi})
\begin{equation}\label{eq:gas1}
\frac{1}{\rho_{\rm gas}}\frac{\dd P}{\dd r}=-\frac{\dd\Phi}{\dd r} = -\frac{GM(r)}{r^2}-\frac{\Upsilon_1 G}{4}\frac{\dd ^2M(r)}{\dd r^2},
\end{equation}
where the dominant contribution to the mass $M(r)$ comes from the dark matter. Assuming spherical symmetry (which is appropriate for these stacked cluster profiles \cite{Wilcox2016}), we have
\begin{equation}\label{eq:masses}
\frac{\dd M}{\dd r}= 4\pi r^2\rho(r)\quad\textrm{and}\quad\frac{\dd^2 M}{\dd r^2}=8\pi r\rho(r)+4\pi r^2\frac{\dd\rho(r)}{\dd r}
\end{equation}
The second term in \eqref{eq:gas1} can be calculated exactly using the NFW profile \eqref{eq:NFW}. The density $\rho_{\rm gas}(r)=\mu m_{\rm p}n_{\rm gas}$ where $m_{\rm p}$ is the proton mass and $\mu$ is the mean molecular weight. One has $n_{\rm gas} = 5n_{\rm e}/(2+\mu)$ where $n_{\rm e}$ is the electron gas density, which we model as an isothermal $\beta$-profile. Using this, one can integrate the pressure support equation \eqref{eq:gas1} to find
\begin{equation}\label{eq:gas2}
P(r) = P(0) - \mu m_{\rm p}\int_0^r n_{\rm e}\left[\frac{G M(r')}{r'^2}+\pi\Upsilon_1 Gr_{\rm s}\rho_{\rm s}\left(1-\frac{r'}{r_{\rm s}}\right)\left(1+\frac{r'}{r_{\rm s}}\right)^3\right]\dd r'.
\end{equation}
This is the beyond Horndeski equivalent of equation 3.2 in \cite{Terukina:2013eqa} and equation 10 in \cite{Wilcox:2015kna}. Note that one can define a thermal mass via \cite{Terukina:2013eqa},
\begin{equation}\label{eq:Mtherm}
M_{\rm therm}(r) = -\frac{r^2}{G\rho(r)}\frac{\dd P}{\dd r}= M(r) +\pi \Upsilon_1r_{\rm s}^3\rho_{\rm s}\left(1-\frac{r}{r_{\rm s}}\right)\left(1+\frac{r}{r_{\rm s}}\right)^3\equiv M(r) + M_1(r),
\end{equation}
where $M_1(r)$ represents the difference between $M_{\rm therm}(r)$ and $M(r)$. So far, we have assumed that the gas is in perfect hydrostatic equilibrium and have ignored any effects of non-thermal pressure. The addition of non-thermal pressure leads to the modified relation $M_{\rm therm} + M_{\rm non-therm}=M + M_1$, where the non-thermal mass comes from the non-thermal pressure $P_{\rm non-therm}$ and the thermal mass is found by solving the hydrostatic equilibrium equation for the thermal part i.e. by assuming an equation of state $P_{\rm therm} = n_{\rm therm} k_{\rm B} T_{\rm therm}$, which is the part deduced from X-ray temperature profiles. \cite{Wilcox:2015kna} found that non-thermal pressure tends to increase the hydrodynamical mass $M_{\rm therm} + M_{\rm non-therm}$ and is therefore degenerate with regions where $M_1>0$, which corresponds to $r<r_{\rm s}$ when $\Upsilon_1>0$ and $r>r_{\rm s}$ when $\Upsilon_1<0$. 

We note that when fitting chameleon gravity models and simulations against our stacked X-ray surface brightness cluster profiles \cite{Wilcox:2015kna,Wilcox2016} we found no evidence for any significant non-thermal pressure component in the outskirts of the clusters (e.g. infall of gas onto the cluster). However, as a precaution we did exclude the central $100$kpcs from these stacked profiles as it is well-established that the centres of clusters are often affected by cooling flows and AGN feedback \cite{brighenti2003feedback}.  

The weak gravitational lensing convergence, caused by light deflection by the cluster mass, is governed by a radial integral along the line of sight of the quantity
\begin{equation}\label{eq:lens1}
\nabla^2(\Phi+\Psi) = 8\pi G\rho_{\rm NFW}(r) .
\end{equation}
Using the fact that $\nabla^2 = r^{-2}\frac{\dd}{\dd r}(r^2\frac{\dd}{\dd r})$ one can multiply \eqref{eq:lens1} by $r^2$ and define a lensing mass
\begin{equation}\label{eq:mWLGR}
M_{\rm WL} = \frac{r^2(\Phi'+\Psi')}{2 G},
\end{equation}
where one has $M_{\rm WL}=M$ in GR. Using equations \eqref{eq:dPhi} and \eqref{eq:dPsi} one can compute the equivalent quantity\footnote{Note that the Jordan frame is used throughout this work and therefore the geodesic equations are unmodified; all of the deviations from GR are captured by equations \eqref{eq:dPhi} and \eqref{eq:dPsi}. } in beyond Horndeski theories
\begin{equation}\label{eq:lens2}
\nabla^2(\Phi+\Psi) = 8\pi G(\rho_{\rm NFW} +\rho_{\rm eff})
\end{equation}
with
\begin{align}
\rho_{\rm eff}(r)&=\frac{1}{8\pi r^2}\frac{\dd}{\dd r}\left(\frac{\Upsilon_1 r^2}{4}\frac{\dd^2M}{\dd r^2}-\frac{5\Upsilon_2r}{4}\frac{\dd M}{\dd r}\right)\nonumber\\&=\frac{\rho_{\rm s}}{4}\left[\Upsilon_1\left(\frac{r_{\rm s}}{r}-2\right)-5\Upsilon_2\left(1+\frac{r_{\rm s}}{r}\right)\right]\left(1+\frac{r}{r_{\rm s}}\right)^{-4}.
\end{align}
One then has
\begin{equation}\label{eq:MWLBH}
M_{\rm WL} = M + \frac{\pi rs^3\rho_{\rm s}}{2}\left[\Upsilon_1\left(\frac{r_{\rm s}}{r}-1\right)-5\Upsilon_2\left(1+\frac{r_{\rm s}}{r}\right)\right]\left(1+\frac{r_{\rm s}}{r}\right)^{-3}\equiv M+M_2,
\end{equation}
where we have again written the contribution from modified gravity as an effective mass showing how the lensing mass differs from the true mass in beyond Horndeski theories.

Ignoring non-thermal pressure, one has the consistency condition $M_{\rm WL}=M_{\rm therm}$ in GR whereas here we have found a different consistency condition $ M_{\rm WL} = M_{\rm therm} - M_1 + M_2 $, which is the generalisation of 2.20 in \cite{Terukina:2013eqa}. This is the essence of our test. Since we can fit jointly for $M_{\rm therm}$ and $ M_{\rm WL}$, any non-agreement (within errors) can be used to place bounds in the $\Upsilon_1$--$\Upsilon_2$ plane. Note from equations \eqref{eq:Mtherm} and \eqref{eq:MWLBH} that it is not possible to tune $\Upsilon_1$ and $\Upsilon_2$ such that $M_{\rm WL}=M_{\rm therm}$, and therefore we do not expect to see any degeneracy between the two beyond Horndeski parameters.

\section{Results}\label{sec:res}

In this section we present the results of our MCMC analysis of lensing and X-ray data, and discuss the implications for beyond Horndeski theories.

\subsection{Constraints}

The data set and methods used to obtain the bounds on $\Upsilon_i$ are identical to those used by \cite{Wilcox:2015kna,Wilcox2016} with the exceptions that the X-ray profiles are now fitted using equation \eqref{eq:gas2} and the convergence is calculated using equation \eqref{eq:lens2}. To fit this model to our lensing profiles we have converted from convergence to shear assuming spherical symmetry \cite{2011ApJ...729..127U}. Briefly, the cluster X-ray surface brightness and lensing profiles are constructed from stacking 58 clusters detected in the XMM Cluster Survey \cite{Romer:1999qt} and Canada France Hawaii Telescope Lensing Survey (CFHTLenS) \cite{2012MNRAS.427..146H}. The CFHTLenS survey analysis combined weak lensing data processing with {\tt THELI} \citep{2013MNRAS.433.2545E}, shear measurement with {\tt lensfit} \citep{2013MNRAS.429.2858M}, and photometric redshift measurement with PSF-matched photometry \citep{2012MNRAS.421.2355H}. These clusters span a range of redshift ($0.1<z<1.2$, median $z=0.33$) and X-ray temperatures ($0.2<T_x<8$ keV, median $T_x=2.3$) and are fully described in \cite{Wilcox:2015kna}. We fit these stacked profiles using a full MCMC analysis (eight parameters covering the NFW parameters $c$ and $M_{200}$, the cluster properties $T_{0}$, $n_{0}$, $b_{1}$, and $r_{1}$, and the modified gravity parameters $\Upsilon_1$ and $\Upsilon_2$) and our methodology has been tested against simulations (both $\Lambda$CDM and chameleon gravity) in \cite{Wilcox2016}. Our MCMC run was a parallelised implementation using 128 walkers with 10000 time steps. We removed the first 2000 iterations as a ``burn in'' phase.

\begin{figure}[h]
\includegraphics[width=0.8\textwidth]{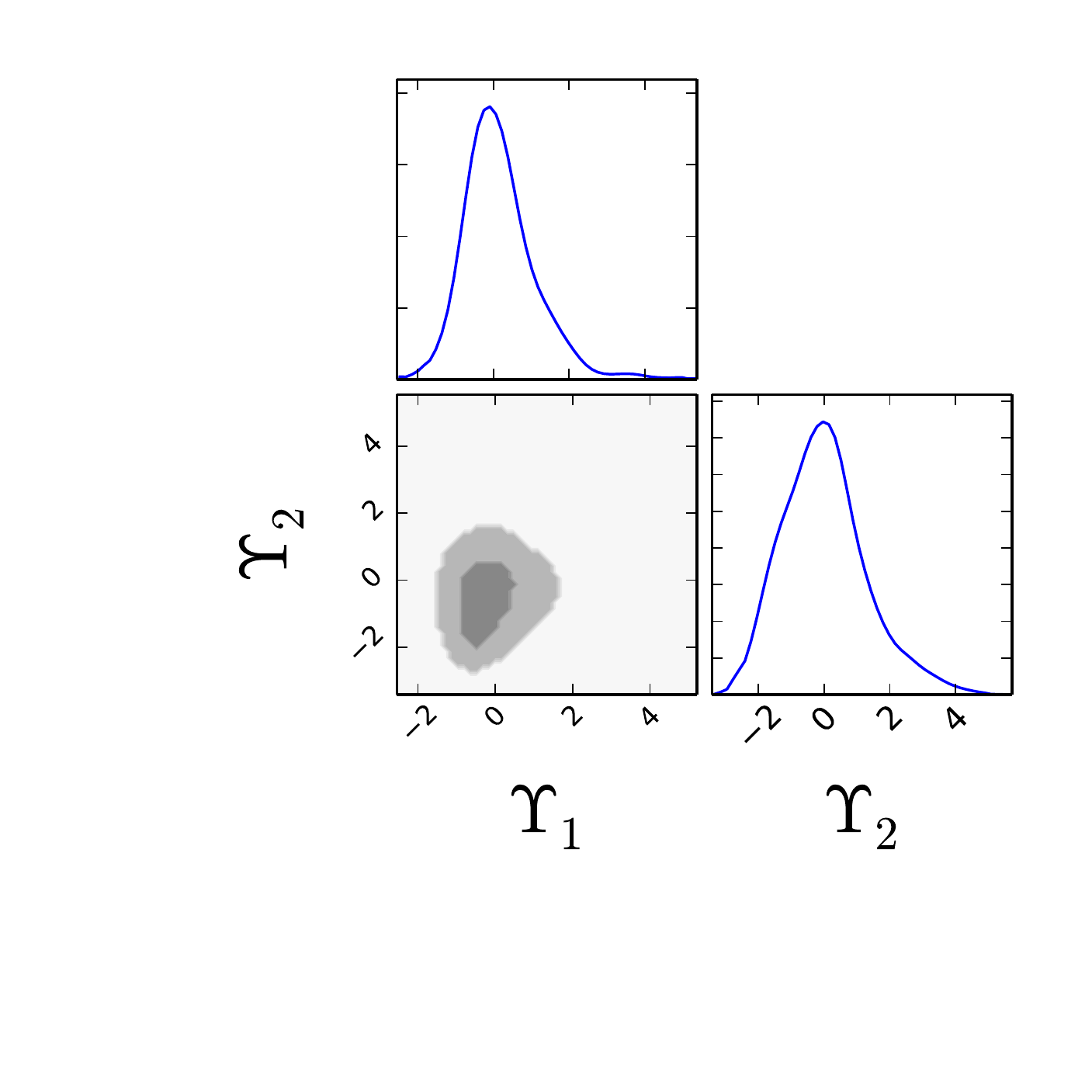}\vspace{-2.7cm}
\caption{The likelihood distributions for $\Upsilon_1$ and $\Upsilon_2$ and the permitted region in the $\Upsilon_1$--$\Upsilon_2$ plane. The dark and light grey contours show the 1- and 2-$\sigma$ allowed regions respectively.}\label{fig:constraints}
\end{figure}

Figure \ref{fig:constraints} shows the results of our MCMC analysis, having marginalised over the other model parameters in \cite{Wilcox:2015kna}. One can see that the data is consistent with GR ($\Upsilon_1=\Upsilon_2=0$) and individually we find
\begin{align}
\Upsilon_{1}=-0.11^{+0.93}_{-0.67}\quad\textrm{and}\quad\Upsilon_{2}=-0.22^{+1.22}_{-1.19},\label{eq:Ucons}
\end{align}
which constitute the best fit values and the 95\% confidence range.  

\subsection{Implications for Beyond Horndeski Theories}

To date, only constraints on $\Upsilon_1$ have been obtained (this is because all of the current constraints come from stellar probes, which do not respond to $\Psi$). The constraint we have derived on $\Upsilon_2$ is the first in the literature. This is important because it provides a second constraint on a different combination of the parameters appearing in the EFT (see equation \eqref{eq:EFT}) and therefore improves our ability to constrain the parameters directly. 

We note that the redshift range of our sample is $0.1<z<1.2$ with median value $z_{\rm med} = 0.33$ and therefore the constraints we have derived here do not apply at redshift zero; they are complimentary to the previous constraints and should be taken to apply at $z_{\rm med}$. Any consistent beyond Horndeski theory should therefore satisfy our constraints at $z=z_{\rm med}$ as well as previous ones that apply at $z=0$. The EFT parameters are typically time-varying and hence so are $\Upsilon_1$ and $\Upsilon_2$. This new constraint at higher redshift therefore restricts the time-dependence of beyond Horndeski cosmologies rather than simply constraining the value of the EFT parameters today. We note that our constraints should be taken somewhat equivocally when applied to models where $\Upsilon_{1,2}$ vary rapidly over the redshift range $0.1<z<1.2$.

\section{Conclusions}\label{sec:concs}

Beyond Horndeski theories represent a very general class of ghost-free scalar-tensor modifications of gravity and therefore encapsulate a wide-variety of dark energy models including galileon theories that admit de-Sitter solutions \cite{DeFelice:2010nf,Babichev:2012re,Kase:2014yya} without the need for a cosmological constant. They are therefore viable alternatives to $\Lambda$CDM\footnote{Note that they do not solve the old cosmological constant problem in the sense that they do not explain why the contribution of the standard model to the vacuum energy does not gravitate.}. On astrophysical scales, the Vainshtein mechanism acts to suppress the scalar force outside of extended bodies but it is partially broken inside (provided there is at least one beyond Horndeski term i.e. $\alpha_H\ne0$), which gives rise to deviations from the Newtonian limit of GR exemplified by equations \eqref{eq:dPhi} and \eqref{eq:dPsi}. This breaking opens up the possibility of distinguishing beyond Horndeski models from GR locally and is completely characterised by two dimensionless parameters $\Upsilon_1$ and $\Upsilon_2$, which parameterise deviations in the $00$- and $ij$- components of the metric ($\Phi$ and $\Psi$) respectively. These parameters are themselves given by combinations of the parameters appearing in the effective field theory of dark energy (see \eqref{eq:EFT}), which characterise the linear cosmology of these theories. Constraints on $\Upsilon_{1,2}$ therefore constitute constraints on the linear cosmology of beyond Horndeski theories.

The modifications have the result that the hydrostatic and lensing mass of clusters are not equal (as they are in GR) but instead differ by an amount which is a function of $\Upsilon_{1,2}$ and the parameters appearing in the NFW profile. In this work, we have exploited this fact by fitting the stacked X-ray and weak lensing profiles of 58 galaxy clusters at median redshift $z_{\rm med}=0.33$. Using a multi-parameter MCMC analysis described above, we have obtained the constraints given in equation \eqref{eq:Ucons}. The constraint on $\Upsilon_2$ is the first to date, and hence allows one to constrain the cosmology of beyond Horndeski theories to tighter levels since it is given by a different combination of EFT parameters to $\Upsilon_1$. Furthermore, our constraints are the first that apply at higher redshifts---previous constraints \cite{Sakstein:2015zoa,Sakstein:2015aac, Jain:2015edg} use stellar effects in the local neighbourhood and hence apply at redshift zero---and therefore represent a constraint not only on the EFT parameters today but also their time-dependence. 

\section*{Acknowledgements}

This work is based on observations obtained with MegaPrime/MegaCam, a joint project of CFHT and CEA/DAPNIA, at the Canada-France-Hawaii Telescope (CFHT) which is operated by the National Research Council (NRC) of Canada, the Institut National des Sciences de l'Univers of the Centre National de la Recherche Scientifique (CNRS) of France, and the University of Hawaii. This research used the facilities of the Canadian Astronomy Data Centre operated by the National Research Council of Canada with the support of the Canadian Space Agency. CFHTLenS data processing was made possible thanks to significant computing support from the NSERC Research Tools and Instruments grantph program.  RN, DB, KK supported by the UK Science and Technology Facilities Council grants ST/K00090X/1.  KK also acknowledges support from the European Research Council grant through 646702 (CosTesGrav).  Numerical computations were performed on the Sciama High Performance Computing (HPC) cluster which is supported by the ICG, SEPNet and the University of Portsmouth.

\bibliographystyle{jhep}
\bibliography{ref}
\appendix

\section{Relation of $\Upsilon_i$ to the EFT Parameters}\label{app:U2}

In this appendix we derive the relations between $\Upsilon_i$ and the parameters appearing in the effective field theory of dark energy. Our goal is not to re-derive the non-relativistic limit, as this was done by \cite{Kobayashi:2014ida,Koyama:2015oma}. Instead, we will follow the procedure of \cite{Saito:2015fza}, and translate the parameterisation used by \cite{Kobayashi:2014ida} into the EFT language. \cite{Kobayashi:2014ida} define the following parameters, which we give in terms of the EFT parameters using the relations given by \cite{Saito:2015fza}\footnote{Note that these relations only apply when cubic terms are absent, and that we have translated the parameters into the mass scale $\Lambda$ defined in \cite{Koyama:2015oma}. The conversion to the scale $\tilde{\Lambda}$ used by \cite{Kobayashi:2014ida,Saito:2015fza} (called $\Lambda$ by \cite{Saito:2015fza}) is $\tilde{\Lambda}^3=\mpl\Lambda^2$. Furthermore, one has $X\rightarrow\mpl^2X$. This means that we work with dimensionless scalars whereas those used by \cite{Kobayashi:2014ida,Saito:2015fza} have mass dimension one.}:
\begin{align}\label{eq:EFTrels}
\alpha_*&=\frac{M_*\Lambda^2}{4\mpl X}\alpha_H\quad\alpha_1=\frac{M_*\Lambda^2}{4\mpl X}(\alpha_H-\alpha_B)\quad\alpha_2= \frac{M_*\Lambda^2}{4\mpl X}\alpha_T\nonumber\\\nu&=\left(\frac{M_*\Lambda^2}{\mpl X}\right)^2\frac{\alpha_H-\alpha_T-\alpha_B}{8}\quad \mathcal{F}=1+\alpha_T\quad\mathcal{G}=1+\alpha_H,
\end{align}
where the speed of tensor perturbations is $c_T^2=1+\alpha_T$. The parameters $\Upsilon_i$ are given by \cite{Kobayashi:2014ida}\footnote{Note that we have made the same choice as \cite{Saito:2015fza} who set $\tilde{M}_{\rm pl}=M_*$ where $\tilde{M}_{\rm pl}$ is defined in \cite{Kobayashi:2014ida}.}
\begin{align}\label{eq:upsfund}
\Upsilon_1&=-8\frac{\alpha_*^2(1+\alpha_B)}{\Xi}\quad\textrm{and} \quad\Upsilon_2=-\frac{8\alpha_*\alpha_1(1+\alpha_B)}{5\Xi},
\end{align}
where
\begin{equation}
\Xi=\mathcal{G}\left(4\alpha_1\alpha_2-2\alpha_1\alpha_*+\mathcal{G}\nu\right)-2\mathcal{F}\alpha_1^2.
\end{equation}
Using the relations \eqref{eq:EFTrels} in \eqref{eq:upsfund} yields equation \eqref{eq:EFT}.
\end{document}